\documentstyle[aps,prl,twocolumn]{revtex}
\input epsf

\begin{document}
\bibliographystyle{prsty}

\draft           %instructs revtex to print out PACS from the \pacs{sdfsdf}
                 %command
                 
%\input psnames  % defines macros for names of PostScript TFMs
%\input pstext   % defines \PStext and \CMtext
%\font\tr=\TimesR at 10pt
%\tr
                                
%--Comment out the following 2 lines to get one-column mode ------------------------
\twocolumn[\hsize\textwidth\columnwidth\hsize\csname
@twocolumnfalse\endcsname
\title{Bulk band gaps in divalent hexaborides}

\author{J. D. Denlinger} 
\address{ Advanced Light Source, Lawrence Berkeley National Laboratory, 
Berkeley CA 94720}
\author{J. A. Clack, J. W. Allen, G.-H. Gweon} 
\address{Randall Laboratory, University of Michigan, Ann Arbor, MI
48109-1120}
\author{D. M. Poirier\cite{dmpaddr}, C. G. Olson}
\address{Ames Laboratory, Iowa State University, Ames, IA 50011}
\author{J. L. Sarrao\cite{lanladdr}, A. D. Bianchi\cite{lanladdr}, 
Z. Fisk}
\address{National High Magnetic Field Lab and Dept. of Physics, 
Florida State University, Tallahassee, FL 32306}
\date{\today}
%\date{September 1, 2000}
\maketitle
%
%
%%%%%%%%%%%%%%%%%%%%%%%%%%%%%%%%%%%%%%%%%%%%%%%%%%%%%%%%%%%%%%%%%%%%%%%%%%%%%%%%%
% ABSTRACT
%$$$$$$$$$$$$$$$$$$$$$$$$$$$$$$$$$$$$$$$$$$$$$$$$$$$$$$$$$$$$$$$$$$$$$$$$$$$$$$$$

%
\begin{abstract}
Complementary angle-resolved photoemission and bulk-sensitive {\it
k-resolved} resonant inelastic x-ray scattering of divalent hexaborides
reveal a $>$1~eV X-point gap between the valence and conduction bands, in
contradiction to the band overlap assumed in several models of their novel
ferromagnetism.  This semiconducting gap implies that carriers detected
in transport measurements arise from defects, and the measured location of
the bulk Fermi level at the bottom of the conduction band implicates boron
vacancies as the origin of the excess electrons. The measured band structure
and X-point gap in CaB$_6$ additionally provide a stringent test case for
proper inclusion of many-body effects in quasi-particle band calculations.

\end{abstract} 

%PACS
%79.60.-i  Photoemission and photoelectron spectra
%71.18.+y  Fermi Surface: calculations and measurements; effective mass, g factor
%71.20.-b  Electron density of states and band structure of crystalline solids
\pacs{PACS numbers:  79.60.-i, 71.18.+y, 71.20.-b}

%--Comment out the following line to get one-column mode ------------------------
\vskip2pc]

%useful definitions here

\def\EF{$E_{\rm F}$}
\def\kF{$k_{\rm F}$}
\def\hv{h$\nu$}
\def\kx{k$_x$}
\def\ky{k$_y$}
\def\kz{k$_z$}
\def\G{$\Gamma$}
\def\sbar{$\overline{\rm S}$}
\def\gbar{$\overline{\Gamma}$}
\def\gradv{$\vec\nabla V$}
\def\kpar{$k_{\parallel}$}
\def\deg{$^{\circ}$}
\def\A-1{$\AA^{-1}$}
\def\B6{B$_6$}
\def\CaLaB6{Ca$_{0.995}$La$_{0.005}$B$_6$}
\def\(1-d){$_{1-\delta}$}
\def\1+d{$_{1+\delta}$}
\def\~{$\approx$}
\def\site{$cite$}

\def\strike#1{% 
{% 
\setbox0=\hbox{#1}% 
\dimen0=\ht0 \dimen1=\dp0 
\setbox1=\vbox{ 
\box0 \vskip-\dimen1\vskip-0.5ex \hrule 
}% 
\ht1=\dimen0 \dp1=\dimen1 
\box1 
}% 
} 

%main text
Great interest in the divalent hexaborides has been generated recently
by the discovery of ferromagnetism (FM) in La-doped Ca\B6\
\cite{Young99} and by exotic theoretical models to explain the unusual
magnetism, e.g. that it represents the ground state of a dilute electron
gas \cite{Ceperley99,Ortiz99} or of a doped excitonic insulator
\cite{Zhitomirsky99,Balents00,Barzykin00,Murakami-condmat02}.  Subsequent
experiments have extended the observation of ferromagnetism also to the
undoped systems of Ca\B6, Sr\B6\ and La-doped Ba\B6\
\cite{Vonlanthen00,Ott00,Terashima00} raising new questions about the
origins of the unusual magnetism.

Central to the excitonic instability model, and indeed the starting
point of most thinking about the divalent hexaborides, is the presumed
existence of small band overlap between the top of the boron
valence states and the bottom of the cation $d$-conduction band at the
X-point of the simple cubic Brillouin zone appropriate to these
materials.  Band overlap is predicted by LDA band structure calculations
\cite{Hasegawa79,Massidda97,Rodriguez00} and de Haas-van Alphen (dHvA)
and Shubnikov-de Haas (SdH)  experiments
\cite{Goodrich98,Aronson99,Hall01} have been interpreted in this
semi-metal framework.  

The well known need for many body corrections to the LDA in calculating
semiconductor band gaps \cite{Hybertson85} calls into question the LDA
band overlap result.  Indeed a recent pseudopotential $GW$ quasiparticle
band calculation for Ca\B6\ has predicted a large 0.8 eV X-point band
gap \cite{Tromp01}. In contrast, two new all-electron $GW$ calculations
have instead predicted an intermediate 0.3 eV bandgap
\cite{Schilfgaarde02}, and an unusual increased band overlap relative to
LDA \cite{Kino-aps02}, the latter thought to be due the special character
of the X-point states. The wide disparity in results for three different
implementations of the $GW$ quasiparticle band calculation scheme, which has
been very successful for calculating band gaps in many common semiconductors
\cite{Hybertson85}, shows clearly that existing methodologies are inconsistent
\cite{Kotani-condmat01,Ku-condmat02}, applied in this case to a system
with which there is no prior experience or firm experimental knowledge.
Thus the question of band overlap versus a band gap (and its magnitude)
is not only  crucial for the novel physics of these materials but also
serves as a particularly pointed test case for one of the most
fundamental aspects of the modern theory of electrons in solids. 

In this paper we present data on divalent hexaborides from angle resolved
photoemission (ARPES) and resonant inelastic x-ray scattering (RIXS)
showing a global $bulk$ electronic structure consistent with the calculated
bulk boron-block band structure, including the existence of a $>$1 eV
X-point band gap that is significantly larger than any of the three $GW$ band
calculations.   For stoichiometric material there are just enough electrons
to fill the boron-block bands so an X-point gap makes the material an
insulator, whereas band overlap makes it a semimetal with hole and electron
Fermi surfaces that enclose equal volumes.  Thus the new band gap model,
first suggested to the community by Ref. 18 and experimentally confirmed
here, requires the reinterpretation of previous bulk sensitive experiments
\cite{Goodrich98,Aronson99,Hall01,Ott97,Fisk79,Degiorgi97} and also the
recognition that the presence of measured metallic carrier densities of
\~$10^{19}$-$10^{20}$ cm$^{-3}$ can only be explained by
off-stoichiometry. Furthermore, Fermi surfaces measured by ARPES for
Sr\B6\ and Eu\B6\ locate the chemical potential in the bottom of the
conduction band and identify the carriers as $n$-type, consistent with the
sign of the Hall coefficient \cite{Fisk79}.  Additionally it is found that
the chemical potential is variable from sample to sample and is also
surface dependent, not surprising for a defect dominated semiconductor.  
The observed excess electrons imply either excess cations or boron
vacancies,  almost certainly the latter since it is not likely that 
excess metal ions can be packed into the rigid \B6\ sublattice. While
such boron vacancies may also produce the magnetic moments
\cite{Monnier-condmat01}, recent detection of Fe impurities
correlated to magnetism suggest an alternate origin
\cite{Taniguchi-condmat02}, and both possibilities suggest some
similarities to dilute magnetic semiconductors \cite{Fisk-sces01}. 

Single crystal samples of Ca\B6, Sr\B6\ and Eu\B6\ were grown from an
aluminum flux using powders prepared by boro-thermally reducing cation
oxides \cite{Ott97}.  ARPES experiments were  performed both at the
undulator beamline 10.0 of the Advanced Light Source synchrotron and at
the Ames/Montana beamline of the Synchrotron Radiation Center (SRC) at
the University of Wisconsin.  Samples oriented by Laue  diffraction were
cleaved in situ to reveal a [100] surface just before the  measurement,
which was done at a sample temperature of 20-30 K and in a vacuum  of
\~4$\times10^{-11}$ Torr.  A photon energy of 30 eV was used to probe
the \G-X band structure, a value internally consistent with an ``inner
potential step'' of 11.2 eV experimentally determined for Eu\B6\ from
photon energy dependent measurements.  The Fermi energy (\EF) and
instrumental resolution were calibrated with a reference spectrum taken
on scraped Au or sputtered Pt foils.  The ALS instrumental resolution
was 22 meV with total angular resolution of 0.3\deg. The SRC
instrumental resolution was 130 meV with an angular resolution of
$\pm$1\deg.   Fermi-energy intensity maps (or ``Fermi surface'' maps)
were acquired by detection of electron emission along two orthogonal
detection angles relative to fixed sample position.  The energy window
for the SRC FS maps was 170 meV.  

Bulk-sensitive soft x-ray emission (SXE) and absorption (XAS)
spectroscopies were performed at the ALS Beamline 8.0 with
%using the Tennessee/Tulane grating spectrometer.  The 
experimental emission and absorption spectral resolutions of \~0.3 eV
and \~0.1 eV, respectively.  SXE, measured with a 1500 lines/mm grating
spectrometer for fixed photon energy excitation at and above the B $1s$
core threshold, is used as a probe of the dipole-selective occupied
valence band (VB) partial density of states (DOS), i.e. boron
p-states.  In the resonant inelastic x-ray scattering (RIXS) regime near
threshold, momentum conservation between valence and conduction electrons
in the final state provides {\it k-resolved} features in the valence SXE
\cite{Ma92}.  XAS, a probe of unoccupied conduction band (CB) states, was
measured both by total electron yield (TEY) as a function of photon
energy and by partial fluorescence yield (PFY) with the detection
window covering the entire VB emission.  
%{\it PFY is preferentially used near threshold because of zero
%pre-threshold background intensities and greater bulk sensitivity than
%TEY.}   

Fig. 1 shows the band structure measured at the ALS for Ca\B6\ along
\G-X compared to the pseudopotential $GW$ band calculation \cite{Tromp01}. 
The ARPES data, shown with reverse grayscale intensities, is the
sum of two data sets with s- and p-polarization geometries that
individually exhibit strong symmetry selection rule effects that we
will present and analyze elsewhere.  All  boron-derived theory bands
(1-6) are easily identified as labeled in Fig. 1 \cite{ARPESnote}.  
The valence band labeled 1 is observed to lie 1.15 eV below \EF while the
theory conduction band 0 is not observed. This indicates that indeed
it is separated by a gap from valence band 1, and that the
theory underestimates the bandgap by at least 0.35 eV.
% {\it since the experimental chemical potential lies either in defect
%states below the conduction band minimum or in the conduction band but so
%near its minimum that any occupation cannot be observed.}  
%{\bf The $GW$ calculation is also observed
%to overestimate the experimental boron bandwidth by \~1.4 eV and
%underestimate the overlap of bands 2 and 3 by \~0.8 eV.}

ARPES data always show the valence band structure of Fig. 1, including
the X-point gap, but often the chemical potential is located in 
conduction band 0 above the gap to create a small X-point electron
pocket.   Fig. 2(a,b) presents SRC data showing only the near \EF\ 
behavior for Sr\B6\ and Eu\B6 with band labeling as in Fig. 1.  Shown in
Fig. 2(c,d) are \kx/\ky\  FS maps of the X-point electron pockets for the
same surfaces  as for the data of Fig. 2(a,b). The maps reveal an
elliptical FS contour  for Eu\B6\ and a smaller, not fully resolved, FS
for Sr\B6.  As defined from peak positions marked in Fig. 2(a), the
observation of the bottom of the conduction band allows a quantitative
measure (1.15$\pm$0.1 eV) of the X-point  gap for Sr\B6, approximately 40\%
larger than the maximum gap found in the  Ca\B6\ $GW$ band calculations. A
similar gap value is inferred for Eu\B6\ in Fig. 2(b), but with larger
uncertainty due to the visible smearing of spectral weight at the top of
band 1. We ascribe this smearing, which occurs also for the other Eu\B6\
boron bands (not shown), but does not occur in Ca\B6\ and Sr\B6\, to the
larger number of B vacancies in Eu\B6\ implied by the larger occupation of
its X-point conduction band.  

SdH and dHvA experiments on Eu\B6 \cite{Goodrich98,Aronson99} also find
ellipsoidal Fermi surfaces, i.e. two frequencies with  the angle dependences
\cite{Aronson99} for the two extremal orbits,  but with four frequencies
total indicating two ellipsoids  having slightly differing sizes.  The
semi-metallic band overlap model identifies these two FS sheets as the
electron and hole pockets while an alternate interpretation, consistent
with our ARPES data, is that a single electron pocket is slightly
spin-split by the large internal field of the ferromagnetically ordered Eu
$4f$ moments,  aided by the very high magnetic field employed in the
measurements \cite{noARPESsplitting}.  Consistent with this
interpretation, dHvA studies of  Ca\B6\ and Sr\B6\ reveal only two
frequencies \cite{Hall01} as expected for a single elliptical conduction
band pocket.  Of the more complex ``lens'' and ``napkin
ring'' FS topologies resulting from band-overlap electron-hole mixing
\cite{Hasegawa79,Rodriguez00}, only the ``lens'' FS has a
possible dHvA correspondence \cite{Hall01} for Ca\B6\ and Sr\B6, and both
are precluded for Eu\B6\ by the observed SdH angular dependences.     

Given the logical necessity for boron vacancies, it is then not
surprising that  considerable variation of the chemical potential
position  from sample to sample and from bulk to surface is a basic
aspect of divalent hexaborides.  For example, the FS dimensions found
for  Eu\B6 by SdH and dHvA \cite{Goodrich98,Aronson99} are  different
both from one another and also from those implied by the  ARPES of Fig.
2.  In one ARPES experiment on Eu\B6\ an X-point  electron pocket was
initially not present and then appeared somewhat abruptly about four
hours after cleavage, but with a size smaller than that of Fig. 2
(a,c).  As expected if boron vacancies are involved, this time dependent
shift of the surface chemical potential was accompanied by some 
redistribution and shift of the boron-block bands, but always  with the
{\it same} band gap value observed previously. Thus, as occurs for other
semiconductors, surface defects and band bending control the surface
chemical potential position.  Elucidating and controlling  the details
of the defect states for both the bulk and the  surface is an essential
goal for future research on these materials.

Bulk-sensitive SXE and XAS {\it quantitatively} confirm the X-point
band structure of Ca\B6\ presented in Fig. 1.   Fig. 3(a) compares
valence emission spectra for Ca\B6\ with at-threshold (187.85 eV) and
above-threshold excitation (212 eV) as indicated by arrows in the TEY
absorption spectrum (Fig. 3 inset).  The above-threshold SXE shows a
lineshape in very good agreement with a calculated boron p-DOS for Sr\B6\
\cite{Massidda97}. The threshold-excited RIXS spectrum, on the other
hand, has a different overall profile and multiple distinct peaks
including a sharp elastic emission peak (labeled e in Fig. 3) at the
threshold onset energy (187.85 eV) as the result of direct radiative decay
from the CB minimum.  This elastic peak provides a convenient marker for
relative calibration of the emission and absorption energy
scales \cite{energy-calib}. The separation of the top of the VB
emission from this absorption-threshold elastic peak is a graphic
signature of the existence of a bulk semiconductor band gap.

Interpreted in the framework of RIXS, the threshold excitation into
the X-point CB minimum results in a superposition of {\it coherent}
momentum-conserving emission and {\it incoherent} emission of
various origins \cite{Ma92}.  As done in previous RIXS studies of
semiconductors \cite{Luning97}, we can refine the RIXS analysis to
highlight the k-selective features if we approximate the
incoherent emission profile by the above-threshold SXE spectrum and
subtract it from the RIXS spectrum with the scaling shown in Fig. 3(a).
The resulting RIXS difference spectrum is shown in Fig. 3(b) with
comparison to the sharpest low binding energy peaks of the X-point Ca\B6\
ARPES spectrum plotted to align peaks 2 and 3.  This alignment produces
excellent agreement between ARPES and RIXS for peak 1, i.e. the VB
maximum.  Furthermore, the PFY absorption edge is also plotted in Fig.
3(b) and it is observed that the ARPES Fermi level occurs at the onset of
the PFY threshold intensity rise.  Thus for Ca\B6, we have provided a bulk
confirmation of all the features of the surface-sensitive ARPES
measurements, in particular, the value of the X-point band gap (1.15 eV)
and the location of the chemical potential just at the bottom of the
conduction band. Soft x-ray measurements of other divalent hexaborides
give similar results \cite{Denlinger-vuv13,Denlinger-unpubl}, and SXE and
ARPES of La-doped Ca\B6\ and trivalent La\B6\ reveal that the gap
magnitude is very sensitive to the conduction band occupation
\cite{Denlinger-unpubl,Mo-condmat01}.   

In summary, complementary ARPES and SXE/XAS experiments on divalent
Ca\B6\ quantitatively agree to provide a clear case for the existence of
a large energy scale bulk band gap. Small energy scale
excitonic-insulator or hybridization gap models based on X-point band
overlap are thus precluded.  Widely differing band gap values from
pseudopotential and all-electron $GW$ calculations illustrate the
insufficiency of current theory, and our high-quality ARPES bulk band
structure of Ca\B6\ sets a clear benchmark for future work.  The
positioning of the bulk chemical potential in the conduction band and
the resulting $n$-type carriers observed in nominally stoichiometric
samples implicate boron vacancies.  The physics of the divalent
hexaborides is not that of an intrinsic semi-metal,  but that of a
defect semiconductor.

\vspace{0.1in}
 
JWA thanks M.C. Aronson for stimulating his interest in 
these materials. We are grateful for very enlightening discussions
with C. Kurdak, O. Gunnarsson and E. Shirley. 
This work was supported at U. of Michigan by the U.S. DoE under
Contract No.  DE-FG02-90ER45416 and by the U.S. NSF Grant No.
DMR-99-71611.  The NHMFL is supported by the U.S. NSF Grant No.
DMR-99-71348. The Ames Lab  is supported by the U.S. DoE under contract
No. W-7405-ENG-82 and the SRC is supported by the U.S. NSF Grant No.
DMR-00-84402. The ALS is supported by the U.S. DoE under Contract No.
DE-AC03-76SF00098.

\vspace{-0.2in}

% REFERENCES

% FIGURE CAPTIONS (for submitted version go here)
%\newpage

% FIGURES

\center{\leavevmode \epsfysize=2.9in\epsfbox{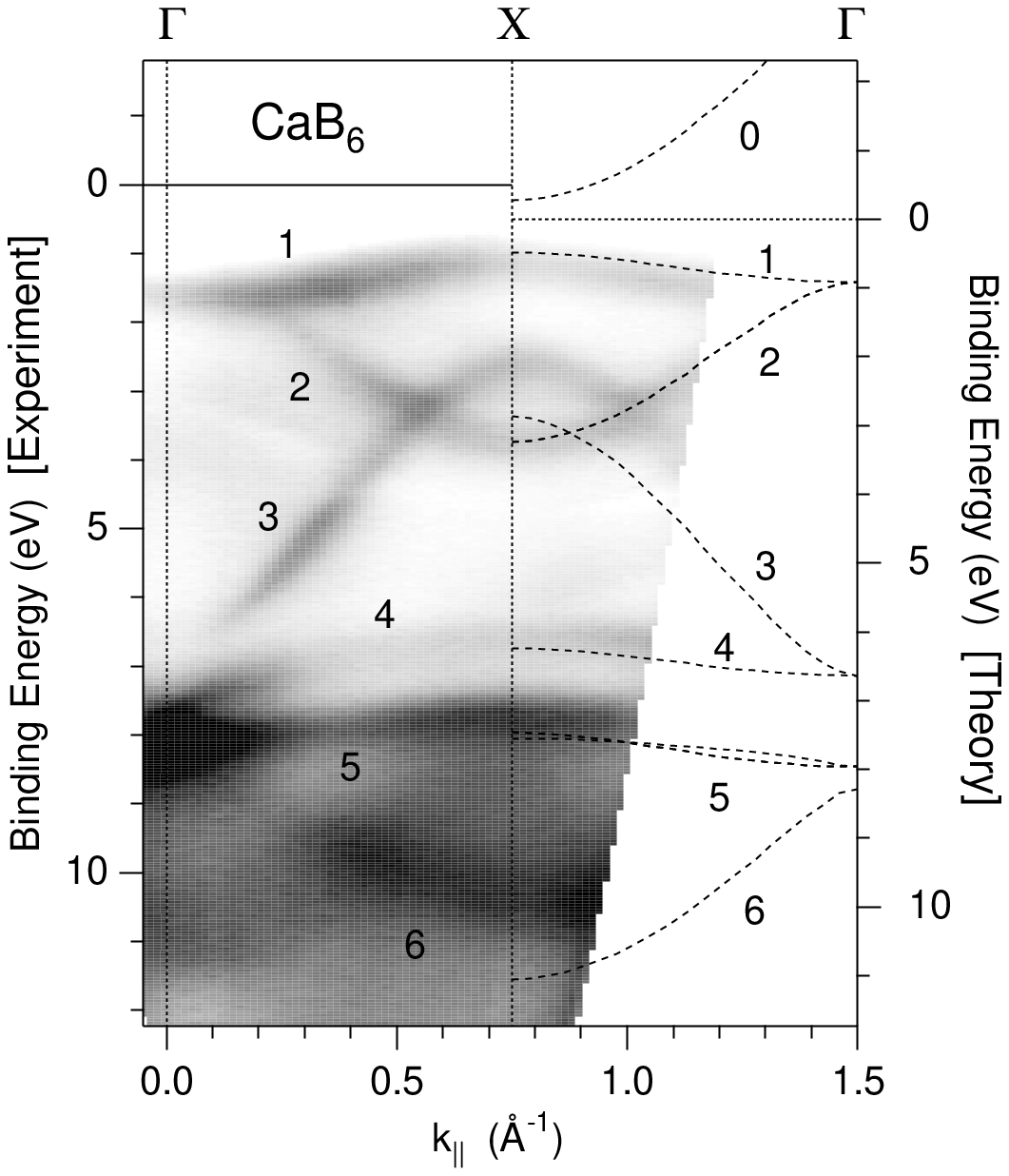}}
\begin{flushleft}
 Fig.\ 1. Comparison of the experimental and theoretical band structures
of Ca\B6\ along \G-X.  The reverse gray scale image of ARPES intensities
is the sum of two data sets with 30 eV s- and p-polarized excitation.
Dashed lines are from the quasiparticle $GW$ calculation\cite{Tromp01}
giving X-point gap between bands 0 and 1.
\end{flushleft}

\center{\leavevmode \epsfysize=2.6in\epsfbox{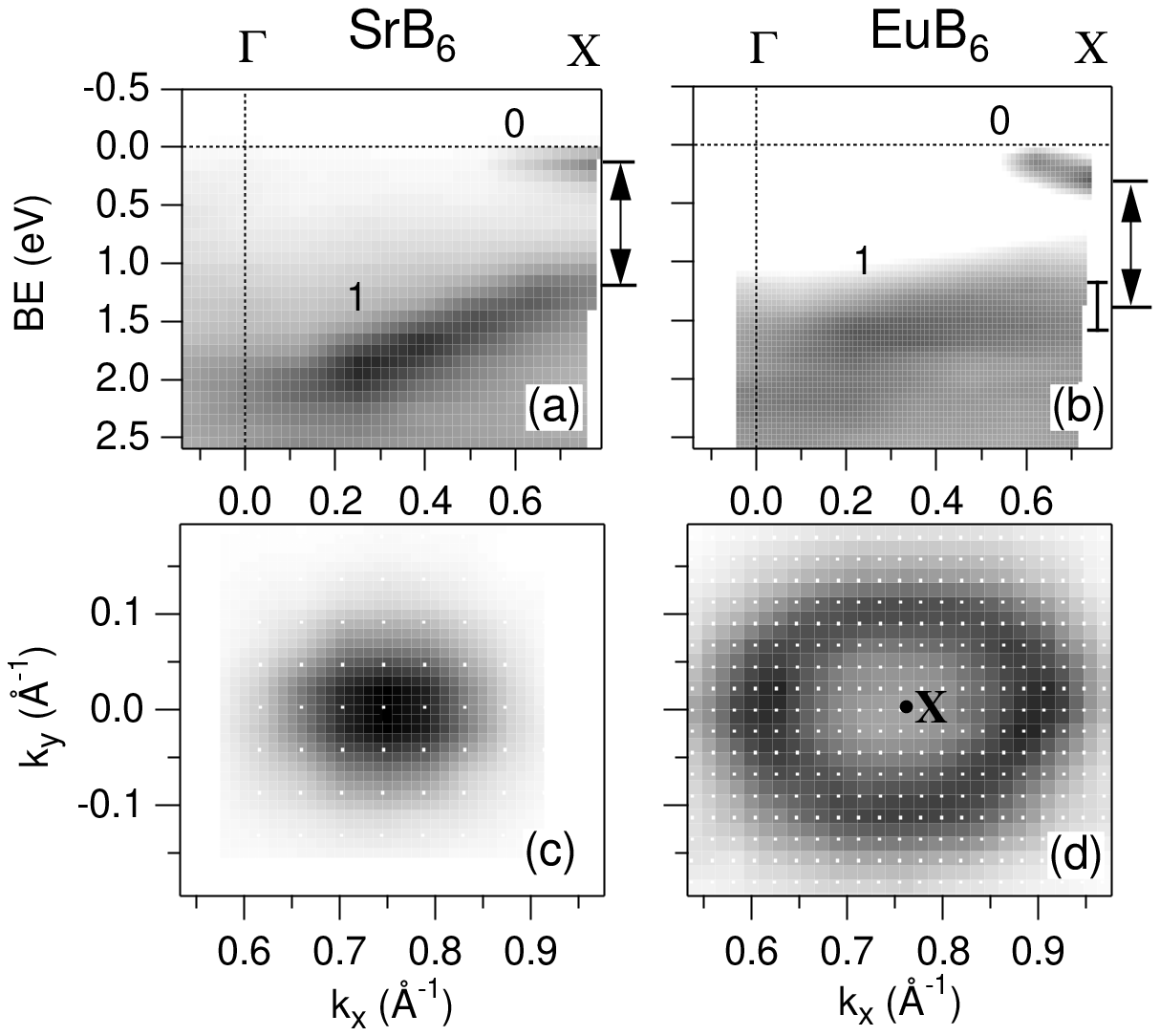}}
\begin{flushleft}
 Fig.\ 2. Near-\EF\ valence band structure for (a) Sr\B6\ and
(b) Eu\B6\ along \G-X showing small band 0 electron pockets above the 
X-point gap to band 1.  \EF\
intensity maps of the X-point at \hv=30 eV for (c) Sr\B6\ and (d) Eu\B6\
showing differing electron pocket sizes.
\end{flushleft}
 
\center{\leavevmode \epsfysize=3.0in\epsfbox{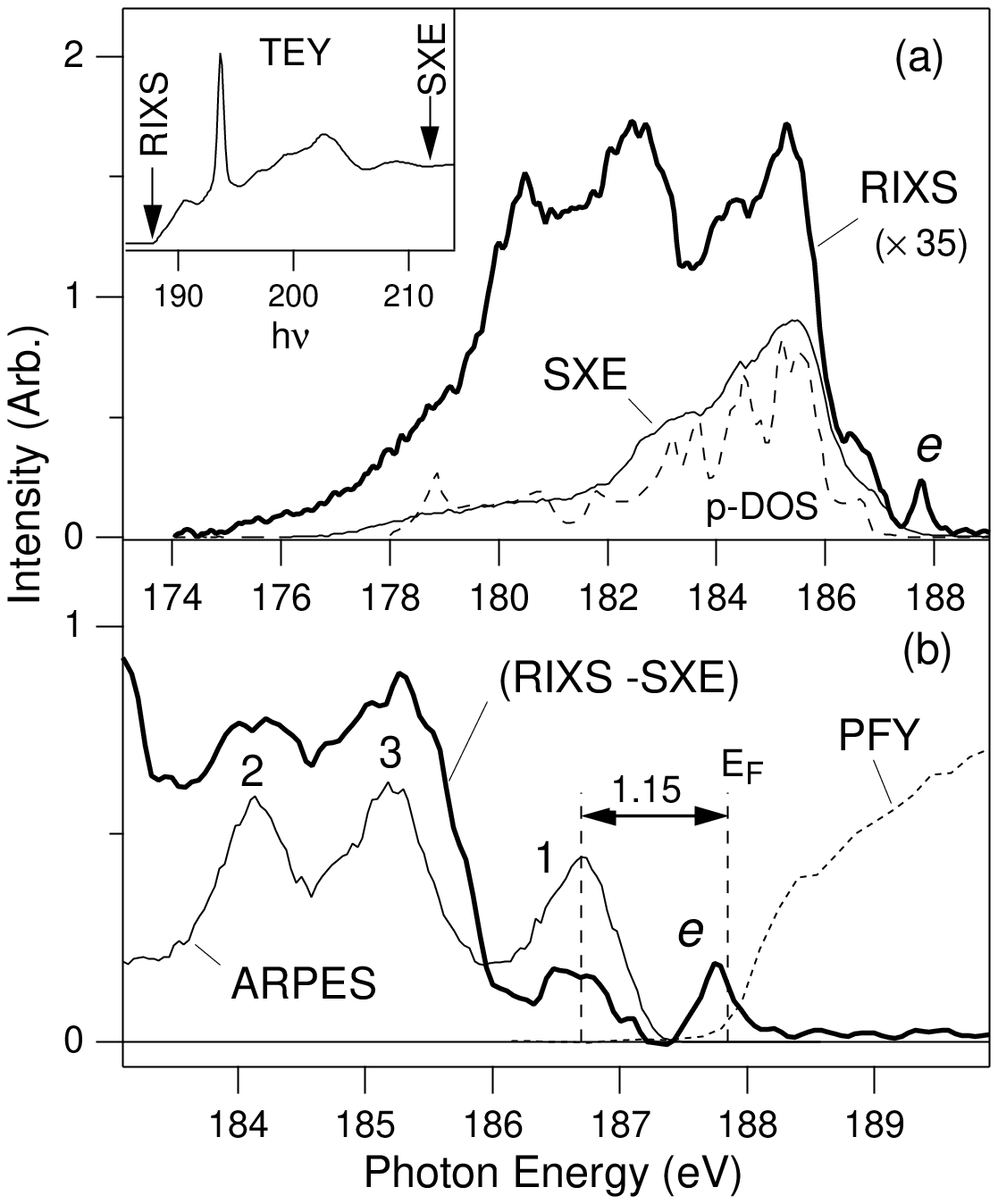}}
\begin{flushleft}
Fig.\ 3.   (a) Ca\B6\ valence x-ray emission excited at-threshold (RIXS)
and above-threshold (SXE) with comparison to calculated
\cite{Massidda97} boron p-DOS (dashed).  (Inset) TEY absorption spectrum
indicating excitation energies.  (b) k-resolved RIXS difference spectrum
(RIXS-SXE) with comparison to ARPES X-point spectrum showing alignment
of valence band peaks and alignment of the ARPES Fermi level to the PFY
absorption threshold.
\end{flushleft}


\begin{thebibliography}{10}
\bibitem{ARPESnote}
An additional weak broad dispersion at the bottom of the valence band,
11 eV at \G, is suggestive of back-folding, perhaps consistent with
2$\times$1 surface order revealed by low energy electron diffraction of
freshly cleaved Ca\B6\ surfaces.
\bibitem{noARPESsplitting}
Our ARPES data are taken in zero field above T${_{Curie}}$ 
%the Curie temperature of
\~15 K and so could not display the splitting.
\bibitem{energy-calib}
Absolute photon energies were calibrated to a 194.0 eV TEY peak due to
surface boron oxide.
\end{thebibliography}
\end{document}